\documentclass[12pt,preprint]{aastex}
\def\jcap{JCAP}
\def\beq{\begin{equation}}
\def\eeq{\end{equation}}
\def\ben{\begin{eqnarray}} 
\def\een{\end{eqnarray}}

\def\dunit{\,h^{-1}\,{\rm Mpc}}
\def\munit{\,h^{-1}\,M_{\odot}}

\def\mth{M_{\rm th}}

\def\jn{{\bf j}_{\rm in}}
\def\em{{\bf e}_{1}}
\def\ei{{\bf e}_{2}}
\def\en{{\bf e}_{3}}

\def\mv{M_{\rm vir}}
\def\lmv{m_{\rm vir}}
\def\rv{r_{\rm vir}}
\def\rth{r_{\rm th}}
\def\rd{r_{\rm d}}
\def\mn{M_{\rm in}}
\def\rn{r_{\rm in}}
\def\rnv{r_{\rm in}/r_{\rm vir}}
\def\rf{r_{f}}
\def\rth{r_{\rm th}}
\usepackage{xcolor}

\begin{document}
\title{The Density Parity Model for the Evolution of the Subhalo Inner Spin Alignments with the Cosmic Web}
\author{Jun-Sung Moon$^{1,2}$ and Jounghun Lee$^{1}$}
\affil{$^1$Astronomy Program, Department of Physics and Astronomy,
Seoul National University, Seoul 08826, Republic of Korea
\email{jsmoon.astro@gmail.com, cosmos.hun@gmail.com}}
\affil{$^2$Research Institute of Basic Sciences, Seoul National University, Seoul 08826, Republic of Korea}
\begin{abstract}
We develop a new model within which the radius-dependent transition of the subhalo inner spins with respect to the 
cosmic web and the variation of the transition threshold radius ($\rth$) with subhalo mass ($\mv$), smoothing scale ($\rf$), and redshift ($z$) 
can be coherently explained. 
The key tenet of this model is that the competition between the pressure effect of the inner mass and the compression effect of the local 
tidal field determines which principal direction of the tidal field the inner spins are aligned with. If the former predominates, 
then only the tidal torques turn on, resulting in the alignments of the inner spins with the intermediate principal axes of the tidal field. 
Otherwise, the subhalo spins acquire a tendency to be aligned with the shortest axes of the subhalo shapes, which is in the major principal  
directions of the tidal field. Quantifying the two effects in terms of the densities, we make a purely analytical prediction for $\rth (\mv, z, \rf)$. 
Testing this model against the numerical results from a high-resolution dark matter only N-body simulation in the redshift range of $0\le z\le 3$ on the 
galactic mass scale of $11.8\le \log \mv/(\munit)\le 12.6$ for two different cases of $\rf/(\dunit)=0.5$ and $1$, 
we find excellent agreements of the model predictions with the numerical results. 
It is also shown that this model naturally predicts the alignments between the inner spins of the present subhalos with the principal axes 
of the high-$z$ tidal field at the progenitors' locations. 
\end{abstract}
\section{Introduction}\label{sec:intro}

The intrinsic spin alignment of subhalos with the cosmic web is a phrase framed to describe the phenomenon that the 
subhalos exhibit preferred directions in their spin orientations with respect to the surrounding anisotropic matter distribution dubbed the cosmic web \citep{web96}. 
As this phenomenon is believed to be closely linked with the acquisition and growth of the subhalo angular momentum as well as with the mechanism responsible for 
the emergence of the cosmic web on the largest scale, it has so far become the subject of numerous theoretical and observational investigations in the field of the 
large-scale structure \citep[see][for a comprehensive review]{review1,review2}.  
What was consonantly revealed by those vigorous studies is that the strength and tendency of the intrinsic spin alignments of subhalos sensitively depend on the redshift, 
subhalo mass, scale and type of the web environment, as well as on the background cosmology 
\citep[e.g.,][]{LP00,pen-etal00,LP01,nav-etal04,bru-etal07,ara-etal07,hah-etal07,LE07,paz-etal08,zha-etal09,cod-etal12,lib-etal13,TL13,for-etal14,pah-etal16,lee-etal18,WK18,gan-etal19,lee-etal20,LL20,wel-etal20,mot-etal21,kra-etal20,dav-etal23}. 

Especially, its variation with subhalo mass has recently drawn deliberative attention. It was shown by multiple N-body simulations that the spin directions of the 
low-mass subhalos tend to be inclined toward the elongated axes of the surrounding filaments, while those of the high-mass counterparts show an opposite inclination of being 
perpendicular to the filaments 
\citep[e.g.,][]{ara-etal07,hah-etal07,paz-etal08,cod-etal12,tro-etal13,lib-etal13,AY14,dub-etal14,for-etal14,cod-etal15a,cod-etal15b,WK17,cod-etal18,gan-etal18,
wan-etal18,gan-etal19,kra-etal20,lee-etal20,LL20,gan-etal21}.  
The mass scale at which the alignment trend is inverted is referred to as the spin transition threshold mass, whose value, $\mth$, at $z=0$ was found in the previous works to be 
approximately in the range of $0.5\le\mth/(10^{12}\munit)\le 5$ \citep[e.g.,][]{ara-etal07,cod-etal12,for-etal14}.  
An approximate empirical formula, $\mth(z)\approx \mth(z=0)(1+z)^{-2.5}$ was suggested by \citet{cod-etal12} to describe the redshift evolution of 
$\mth$ in the framework of the {\it conditional tidal torque theory} \citep[see also][]{cod-etal15b}. 

It was, however, pointed out by several authors that the value of $\mth$ sensitively varies not only with the redshift but also with the type, scale, and thickness of the cosmic web 
\citep[e.g.,][]{for-etal14,gan-etal18,lee-etal21}. 
In other words, the previous estimates of $\mth (z)$ turned out not to be robust against the alteration of how to define the cosmic web, suffering from a certain degree of 
ambiguity. To minimize this ambiguity and to accommodate the environmental variation of $\mth$, 
\citet{lee-etal20} devised a sophisticated algorithm based on the Kolmogorov–Smirnov (KS) test, which can be applied to all four types and scales of the web environments. 
Basically, \citet{lee-etal20} used the principal axes of the local tidal fields, so-called the Tweb \citep{for-etal14,lib-etal18}, and 
refined the definition of the transition threshold as the mass range where the KS test rejects at the confidence level lower than $99.9\%$ the null hypothesis that the probability 
density of an angle of the subhalo spin axis relative to the Tweb intermediate principal axis has the same distribution as that with the Tweb minor principal axis. 
With the help of this algorithm, they were able to coherently depict how $\mth$ changes with the smoothing scale, redshift, and cosmic web type. 
Furthermore, this new algorithm led to the discovery that the value of $\mth$ also depends sensitively on the initial conditions. \citet{lee-etal20} applied this algorithm to the 
halo catalogs from the Cosmological Massive Neutrino Simulations \citep{liu-etal18} and showed that the presence of massive neutrinos has a significant effect of 
lowering the value of $\mth$. \citet{LL20} analyzed the data from the Dark Energy Universe Simulation \citep{ras-etal10,ali-etal12,bou-etal15} with this algorithm and 
showed that the transition threshold has a potential to discriminate dynamical dark energy models from the standard paradigm where the cosmological constant 
$\Lambda$ and cold dark matter (CDM) dominate the present energy density. 

Upon this discovery, a critical question arose whether or not the observable stellar spins of subhalos show similar mass-dependent transitions. 
To answer this question, \citet{lee-etal21} investigated the stellar spin alignments of the subhalos with the Tweb principal axes from a high-resolution hydrodynamical simulation 
and found a somewhat surprising result that the subhalo stellar spins exhibit a peculiar transition between the Tweb intermediate and {\it major} principal axes, while 
their DM counterparts are always perpendicular to the Tweb major principal axes. Noting, however, that in the analysis of \citet{lee-etal21} the subhalo stellar spins were 
measured at twice the half stellar mass radii, much smaller than the virial radii at which the DM spins are usually measured, \citet{ML22} suspected that this difference might be 
responsible for the peculiar alignments of the subhalo stellar spins with the Tweb major principal axes. 
Utilizing the data from a DM-only simulation, they investigated how the DM spins change their alignment tendency as a function of radial distances and proved that the DM spins, 
if measured at much inner radii like the stellar counterparts, exhibited the same peculiar transition between the Tweb intermediate and major principal axes. 

In this paper, we aim at constructing a physical model within which this {\it radius-dependent transition} of the subhalo inner spins can be analytically described. 
For this aim, it should be of vital importance to numerically explore how the alignment tendency of the subhalo inner spins with the Tweb principal axes evolves with redshifts. 
The organization of this paper is as follows. 
Section \ref{sec:num} provides a detailed description of the numerical measurements of the subhalo inner spins and the Tweb principal directions from a simulation dataset. 
Sections \ref{sec:rthz} and \ref{sec:crz} present numerically obtained results on the redshift evolution of the radius-dependent spin transition threshold and on the alignments 
between the inner spins of the present subhalos and the Tweb principal directions at the progenitor redshifts, respectively. 
Section \ref{sec:ana} presents an analytic model for the radius-dependent spin transition phenomenon and a comparison of its predictions with the numerical results.
Section \ref{sec:sum} is devoted to summarizing the main achievements and to discussing their physical implication as well. 

\section{Numerical Analysis}

\subsection{Simulation Data}\label{sec:num}

Our analysis utilizes the subhalo catalogs from the cosmological TNG300-1-Dark simulation that was subsumed 
under the IllustrisTNG project \citep{tngintro1, tngintro2, tngintro3, tngintro4, tngintro5, illustris19} for a $\Lambda$CDM cosmology \citep{planck16}. 
The TNG300-1-Dark simulation box has a side length of $205\,h^{-1}$Mpc, containing $2500^{3}$ DM particles with mass resolution of $4.727\times 10^{7}\,\munit$. 
The subhalos were identified via the Subfind algorithm \citep{subfind} as the substructures of the friends-of-friends (FoF) groups at various redshifts 
in the range of $0\le z\le 20$. A detailed description of the TNG300-1-Dark simulation is provided in the IllustrisTNG web page\footnote{https://www.tng-project.org}. 

We will consider the subhalos identified at redshifts $z \le 3$ with total mass $\mv$ in the logarithmic range $11.8\le \lmv\equiv\log \mv/(\munit) \le12.6$,   
where $\mv$ corresponds to the sum of the masses of all constituent particles. Following the conventional scheme, we determine 
the virial radius, $\rv$, of each subhalo at a given redshift as the radial distance from the subhalo center (i.e., potential minimum) within which the spherically 
averaged mass density contrast reaches $200$ times the critical density, $\rho_{\rm c}(z)$, of the universe.  

With the same routine delineated in our prior work \citep{ML22}, we construct a tidal field, ${\bf T}\equiv (T_{ij})$ smoothed with a Gaussian filter with a scale radius 
of $\rf$, from the TNG300-1 snapshots on $512^{3}$ grids, and define the Tweb configuration surrounding a selected subhalo in terms of the principal 
directions of ${\bf T}$ \citep{for-etal14}. 
A similarity transformation of ${\bf T}$ at the location of each subhalo yields a set of its three orthonormal eigenvectors, $\{\em,\ei,\en\}$, 
corresponding to three eigenvalues, $\{\lambda_{1},\lambda_{2},\lambda_{3}\vert \lambda_{1}\ge\lambda_{2}\ge\lambda_{3}\}$, as the Tweb major, intermediate 
and minor principal axes \citep[see also][]{lee-etal22}. 
The directions of the maximum and minimum matter compression around each subhalo are parallel to $\em$ and $\en$, respectively, while the direction of the 
tidal torquing occurs preferentially along $\ei$ \citep{LP00}. 

Using the information on the comoving positions and peculiar velocities of the constituent DM particles,  we determine the inner angular momentum vectors, ${\bf J}(\rn)$, 
of each subhalo at an inner radius, $\rn$ as 
\begin{equation}
\label{eqn:angm}
{\bf J} (\rn) = \sum_{\alpha=1}^{n_{p}}{m_{p}}\,[({\bf x}_{\alpha}-{\bf x}_{c})\times({\bf v}_{\alpha}-{\bf v}_{c})]\, ,\\
\end{equation}
where $n_{p}$ is the number of the constituent DM particles within $\rn$, ${\bf x}_{\alpha} = (x_{\alpha\,i})$ and ${\bf v}_{\alpha}= (v_{\alpha\,i})$ are the comoving position 
and peculiar velocity of the $\alpha$th particle within $\rn$,  respectively, while ${\bf x}_{c}=(x_{c,i})$ and ${\bf v}_{c}= (v_{c,i})$ are the comoving position and peculiar 
velocity of the subhalo center, respectively. 
Hereafter, the unit inner angular momentum vectors are denoted by $\jn \equiv {\bf J} (\rn) / |{\bf J} (\rn)|$.

\subsection{Evolution of the Threshold Radius for Spin Transition}\label{sec:rthz}

We take the following steps to numerically determine the evolution of the radius-dependent spin transition threshold, $\rth(z)$, over $0\le z\le 3$:

\begin{enumerate}
\item
For each selected subhalo embedded in a Tweb on the scale of $\rf=0.5\dunit$ at a given redshift, we measure the alignments between its inner spin and the 
Tweb principal axes, $\{\vert\jn(z)\cdot{\bf e}_{i}(z)\vert\}_{i=1}^{3}$, at $10$ different points of $\rnv\equiv\kappa/10$, where the integer number $\kappa$ 
varies from $1$ to $10$.
\item
Splitting the logarithmic mass range, $11.8\le\lmv\le 12.6$, into four short intervals each of which has an equal length of $\Delta\lmv = 0.2$, we take the ensemble averages, 
$\{\langle\vert\jn(z)\cdot{\bf e}_{i}(z)\vert\rangle\}_{i=1}^{3}$, separately over the subhalos whose values of $\lmv$ fall in each interval. 
For the evaluation of this statistical average at a given $\rnv$, we constantly apply the DM particle number cut of $n_{p}> 300$, excluding those subhalos 
with $n_{p}(\rnv)\le300$ at the smallest $\rnv$ to avoid any false signal of alignment produced by the numerical flukes \citep{bet-etal07}, and calculate the associated error, 
$\sigma_{i}$, via the Jackknife resampling method\footnote{We utilize the python subroutines included in the astropy package distributed by \citet{astro_pack22}.}.
\item
We employ the transition threshold finding algorithm that was originally developed by \citet{lee-etal20} to find the mass-dependent transition threshold.  
In accordance with this algorithm the transition threshold, $\rth$, can be defined as the inner radius where the null hypothesis of $p(\cos\theta_{1})=p(\cos\theta_{2})$ is rejected 
by the KS test at the confidence level lower than $99.9\%$, where $p(\cos\theta_{1})$ and $p(\cos\theta_{2})$ denote the probability 
densities of $\cos\theta_{1}\equiv\vert\jn(z)\cdot{\bf e}_{1}(z)\vert$ and $\cos\theta_{2}\equiv\vert\jn(z)\cdot{\bf e}_{2}(z)\vert$, respectively. 
\item
Measuring the maximum distance, $D_{\rm max}$, between the cumulative probabilities,  
$P(\ge\!\cos\theta_{1})\equiv \int_{\cos\theta_{1}}^{\infty}d\cos\theta^{\prime}_{1}\,p(\cos\theta_{1}^{\prime})$ 
and $P(\ge\!\cos\theta_{2})\equiv\int_{\cos\theta_{2}}^{\infty}d\cos\theta^{\prime}_{2}\,p(\cos\theta_{2}^{\prime})$ at each $\rnv$ 
from the $N_{\rm in}$ subhalos belonging to a given $\lmv$ interval, we basically locate the $\rnv$ where $\tilde{D}_{\rm max}(\rnv)<1.949$. 
with $\tilde{D}_{\rm max}\equiv \sqrt{N_{\rm in}/2}D_{\rm max}$ and determine it as $\rth/\rv$. The associated errors, $\sigma_{D}$, are also calculated as 
one standard deviation estimated from the Jackknife resamples. The critical value $1.949$ corresponds to the confidence level of $99.9\%$.
\item
Repeating the whole process at various different redshifts in the range of $0\le z\le 3$, we determine $\rth (z)/\rv (z)$. We also test the robustness of the resulting $\rth(z)/\rv(z)$ 
against the variation of $r_{f}$.  
\end{enumerate}

Figure \ref{fig:cl005} plots the inner spin alignments with the Tweb principal axes (color filled symbols) and the results of the KS test of the null hypothesis 
as a function of $\rnv$ (black filled circles) in the left and right columns, respectively, for the case of $\rf=0.5\dunit$ at $z=0$. The four rows correspond to the four different 
$\lmv$-intervals. In the right columns, the dashed lines correspond to the case that the null hypothesis of $p(\cos\theta_{1})=p(\cos\theta_{2})$ is rejected 
at the confidence level $99.9\%$. 
Figures \ref{fig:cl105}--\ref{fig:cl205} plot the same as Figure \ref{fig:cl005} but at $z=1$ and $2$, respectively, while Figures \ref{fig:cl010}--\ref{fig:cl210} plot 
the same as Figures \ref{fig:cl005}--\ref{fig:cl205}, respectively, but for the case of $\rf=1\dunit$. 
As can be seen, the algorithm based on the KS test indeed captures well the threshold inner radius where the preferred spin directions transit from the Tweb intermediate 
to major principal axes as $\rnv$ decreases. 

Figures \ref{fig:z05}--\ref{fig:z10} show the transition threshold radius zone determined by the aforementioned algorithm as a function of redshift 
from the subhalos belonging to the four $\lmv$-intervals for the cases of $\rf/\dunit=0.5$ and $1$, respectively. 
The red filled circles corresponds to the inner radius where $\tilde{D}_{\rm max}$ has its minimum, and the error bar represents the threshold radius zone 
where the null hypothesis is rejected at the confidence level lower than $99.9\%$, i.e., $\tilde{D}_{\rm max}(\rnv)<1.949$. As can be seen, the transition 
threshold radius zone does not evolve sensitively with redshifts for all of the cases of $\lmv$ and $\rf$ considered, in this case that the subhalo inner spins and 
the Tweb principal axes are determined at the same redshifts.

\subsection{Alignments of the Present Inner Spins with the High-$z$ Tidal Fields}\label{sec:crz}

Now that we have pulled it off to trace the redshift evolution of the inner spin alignments with the Tweb principal axes, we would like to investigate how much memory 
the inner spins of the present subhalos at $z=0$ retain about the tidal interactions that their progenitors experience at $z>0$. It is naturally expected that 
the memory of the earlier tidal interactions is better retained by the inner spins than the virial counterparts since the latter is much more susceptible to the 
subsequent infall and accretion processes which play the role of diluting away the earlier memory.

We track the trajectories of the main progenitors\footnote{It is based on the merger trees created by the Sublink algorithm. The main progenitors are defined along the tree 
branches having the `most massive history' \citep{rod-etal15}.} of the present subhalos to higher redshifts and find their locations at earlier epochs. Then, following the same 
procedure described in Section \ref{sec:rthz}, we measure the alignments between the inner spins of the present subhalos and the principal directions of the high-$z$ Tweb 
at the location of their main progenitors, $\{\langle\vert\jn(z=0)\cdot{\bf e}_{i}(z>0)\vert\rangle\}_{i=1}^{3}$, and determine the radius-dependent spin transition threshold, 
$\rth/\rv$, as a function of the progenitor redshifts, $z>0$. 
Figures \ref{fig:cr105}--\ref{fig:cr205} (Figures \ref{fig:cr110}--\ref{fig:cr210}) plot the same as Figures \ref{fig:cl105}--\ref{fig:cl205} 
(Figures \ref{fig:cl110}--\ref{fig:cl210}), respectively, but for the case that $\jn$ are measured at the present epochs while $z$ denotes the progenitor epochs. 
As can be seen, the present inner spins indeed retain quite well the memory of the high-$z$ Tweb, showing a clear signal of the 
radius-dependent transitions in their preferred directions between the intermediate and major principal axes of the high-$z$ Tweb at the locations of the 
main progenitors.  

It is somewhat surprising to witness from all of the $\lmv$ intervals that the inner spin transition occurs at larger inner radii when the inner spins are measured at the 
present epoch rather than at the same high redshifts as the Tweb is constructed. 
The major principal axes of the high-$z$ Tweb at the progenitor locations are much more strongly aligned with the present inner spins than with the 
high-$z$ inner spins, while the intermediate principal axes of the high-$z$ Tweb are more strongly aligned with the same high-$z$ virial spins than with the 
present counterparts. The alignments between the present inner spins and the Tweb major principal axes become stronger when the Tweb is smoothed on a smaller 
scale and when the progenitors are located at higher redshifts. Note that for the case of $\rf=0.5\dunit$ and $z=2$, the present subhalos in the $\lmv$ interval of 
$[12.4,12.6]$ yield no signal of the inner spin transition. Their inner (and virial) spins are always aligned with the {\it major} principal axes of the high-$z$ Tweb 
at the progenitor locations, regardless of the inner radii. 
Figures \ref{fig:zr05}--\ref{fig:zr10} plot the same as Figures \ref{fig:z05}--\ref{fig:z10} but as a function of the progenitor redshifts. Compared with the results 
shown in Figures \ref{fig:z05}--\ref{fig:z10}, the transition threshold radius evolves much more rapidly with the progenitor redshifts. 

To examine whether or not the present inner spins are aligned with the high-$z$ Tweb only for the case of the main progenitors, we repeat the same process but 
with the Tweb at the locations of the non-main progenitors of the present subhalos tracked back to $z=1$ and $2$, the results of which are shown in the left and right panels of 
Figure \ref{fig:cr05_non}, respectively. Regarding the non-main progenitors, we consider those progenitors whose masses are higher than one-tenth of the most massive progenitor's mass 
at each redshift but not the main one. As can be seen, a similar alignment tendency between $\jn(z=0)$ and ${\bf e}_{1}(z>0)$ exists, regardless of the 
progenitor mass. 

We also examine whether the signal of the alignments between $\jn(z=0)$ and ${\bf e}_{1}(z>0)$ is real or spurious by repeating the whole process but with the 
high-$z$ Tweb measured at the shuffled locations of the main progenitors. The left and right panels of Figure \ref{fig:cr05_ran} plot the same as the left panels 
of Figures \ref{fig:cr105}--\ref{fig:cr205}, respectively, but for the case that the locations of the main progenitors are randomly shuffled. As can be seen, no alignment 
signal is found between the present subhalo inner spins and the high-$z$ Tweb principal axes if the progenitor locations are wrongly matched, which verifies that the 
signals shown in Figures \ref{fig:cr105}--\ref{fig:zr10} are real.

\section{Analytic Prediction}\label{sec:ana}

To physically explain the numerical results laid out in Sections \ref{sec:rthz}--\ref{sec:crz}, we put forth the following scenario, calling it {\it the density parity model}.
The external tidal field smoothed on the scale of $\rf$ has two-fold effects on a subhalo: One is to exert a tidal torque, aligning the subhalo spin axis with 
the Tweb intermediate principal direction as predicted by the linear tidal torque theory \citep{whi84,LP00}. The other is to exert a tidal compression, 
aligning the shortest axis of a subhalo shape with the Tweb major principal axis. 
If the latter effect predominates over the former, then the subhalo spin axis also acquires a tendency of being aligned with the Tweb major principal axis, since the 
subhalo spin axis tends to be aligned with the shortest axis of its shape under effective tidal compression \citep[e.g.,][]{LM22}. 
The latter effect, however, can turn on only when it is stronger than the subhalo internal tension.  In other words, if the tension of the interior matter distribution inside 
a subhalo is high enough to resist the external tidal compression, then the subhalo becomes subject only to the tidal torquing but not to the compression, having its spin 
aligned with the Tweb intermediate principal axis. 

Quantifying the counter-balance between the internal tension and tidal compression in terms of densities, we claim that the condition for the occurrence of the 
inner spin transition between the Tweb intermediate and major principal axes at redshift $z$ is nothing but
\begin{equation}
\label{eqn:th}
\rho(r=\rv;M=\mn,z)=\rho(r=\rf;M=\mv,z)\, ,
\end{equation}
where $\rho(r=\rv;M=\mn,z)$ is density contributed by the mass $\mn$ enclosed by the subhalo inner radius $\rn$ at a distance equivalent to $\rv$ and 
$\rho(r=\rf;M=\mv,z)$ is the density around a subhalo of virial mass $\mv$ at a distance equivalent to the comoving smoothing scale $\rf$. 
The former is a measure of the strength of the tidal compression, while the latter quantifies the internal outward tension exerted by $\mn$.  
If $\rho(r=\rv;M=\mn,z)>\rho(r=\rf;M=\mv,z)$,  then the inner spin measured at $\rn$ would be aligned with the Tweb intermediate principal axis 
as the internal tension resists the tidal compression. Otherwise,  the inner spin measured at $\rn$ would be aligned with the Tweb major principal axis 
since the tidal compression wins over the internal pressure.

To compute the density around an overdense region of mass $M$ enclosed by a radius $r$ at a distance $\rd$, we employ the analytic formula that \citet{pra-etal06} 
derived as a halo matter density profile and proved to be valid even beyond the halo virial radii:
\begin{equation}
\label{eqn:dprofile}
\rho\left[\rd;M(r),z\right] = \rho_{c}(z)\delta_{s}(M,z)\exp\left[-2n\left(\frac{r_{d}c}{r}\right)^{1/n}+2n\right]+ \bar{\rho}_{m}(z)\, .
\end{equation}
Here, $\bar{\rho}_{m}(z)$ is the mean matter density, $n$ is the S$\acute{\rm e}$rsic index ranging from $6$ to $7.5$,  $\rho_{c}(z)\equiv 3H(z)^{2}/8\pi G$ is the critical density 
of the universe at $z$ with Hubble parameter $H(z)$, $c$ is the concentration parameter,  and $\delta_{s}$ is the characteristic density contrast related to $c$ 
as \citep{lud-etal14}
\begin{equation}
\label{eqn:dels}
\delta_{s} = \frac{200c^{3}}{3}\left[\ln (1+c) - \frac{c}{1+c}\right]^{-1}\, .
\end{equation}
As an extended version of the standard Navarro-Frenk-White (NFW) density profile \citep{nfw96,nfw97}, Equation (\ref{eqn:dprofile}) has been shown to 
successfully describe the density profiles inside and outside of the galactic halos in the mass range of $10^{11}\le M/(\munit)\le 5\times 10^{12}$ \citep{pra-etal06}.   
For the concentration parameter, $c=c(M,z)$, we also adopt the following empirical relation given by \citet{pra-etal12},
\begin{equation}
\label{eqn:cM}
c(M,z) = \frac{9.7}{1+z}\left[\frac{M}{10^{12}\,h^{-1}M_{\odot}}\right]^{-0.074}\, , 
\end{equation}
which was proven to be valid over a wide mass range of $5\times 10^{10}\le M/(\munit)\le 10^{15}$.

The aforementioned condition for the occurrence of the subhalo spin transition can now be expressed as
\begin{equation}
\label{eqn:thsol}
\delta_{\rm s,in}\exp\left[-2n\left(\frac{r_{\rm vir}c_{\rm in}}{r_{\rm in}}\right)^{1/n}\right]=\delta_{\rm s,vir}\exp\left[-2n\left(\frac{r_{f}c_{\rm vir}}{r_{\rm vir}}\right)^{1/n}\right]\, ,
\end{equation}
where $c_{\rm in}\equiv c(M_{\rm in},z)$ and $c_{\rm vir}\equiv c(\mv,z)$, and $\delta_{\rm s,in}\equiv\delta_{s}(c_{\rm in},z)$,  $\delta_{\rm s,vir}\equiv\delta_{s}(c_{\rm vir},z)$. 
Solving Equation (\ref{eqn:thsol}) about $\rn$, one can find the threshold radius, $\rn=\rth$, for the spin transition between the Tweb major and intermediate 
principal axes at redshift $z$. 
Figures \ref{fig:z05}--\ref{fig:z10} plots the analytical prediction of this density parity model for $\rth(z)$ as gray area for the cases of $\rf/(\dunit)=0.5$ and $1$, respectively.  
For the analytical evaluation of $\rth(z)$, we set the S$\acute{\rm e}$rsic index in Equation (\ref{eqn:dprofile}) at $n=6.0$, the best-fit value on the 
galactic scale \citep{pra-etal06}. 
In a given $\lmv$-interval, we put the maximum and minimum values of $\mv$ and corresponding $\rv$ into Equation (\ref{eqn:thsol}) and then solve it for each case to 
determine the lower and upper limits on $\rth$, respectively. The values of $\rth$ bounded by the upper and lower limits are shown as the gray area in each panel of 
Figures \ref{fig:z05}--\ref{fig:z10}. As can be seen, the analytic predictions describe fairly well the overall trend of $\rth(z)$ for all of the cases of $\lmv$ and $\rf$.

Our density parity model is also capable of predicting $\rth(z)$ as a function of progenitor redshift, $z$. 
For the case that the subhalo spins are measured at $z=0$ while the Tweb principal axes are measured at the progenitor locations tracked to $z>0$, 
the spin transition threshold radius can be evaluated by equating $\rho(r_{\rm vir,0};M_{\rm in,0},z=0)$ to $\rho(\rf(z);M=\mv,z>0)$: 
\begin{equation}
\label{eqn:thsol2}
\rho_{c0}\delta_{\rm s,in,0}\exp\left[-2n\left(\frac{r_{\rm vir,0}c_{\rm in,0}}{r_{\rm in,0}}\right)^{\frac{1}{n}}+2n\right] + \bar{\rho}_{m0} = 
\rho_{c}(z)\delta_{\rm s,vir}\exp\left[-2n\left(\frac{r_{f}(z)c_{\rm vir}}{r_{\rm vir}}\right)^{\frac{1}{n}}+2n\right] + \bar{\rho}_{m}\, ,
\end{equation}
where $\bar{\rho}_{m0}=\bar{\rho}_{m}(z=0)$, $\rho_{c0}=\rho_{c}(z=0)$, $\delta_{\rm s,in,0}=\delta_{s}(M=\mn,z=0)$, 
$M_{\rm vir,0}=(4\pi/3)\Delta_{c}\rho_{c0}r^{3}_{\rm vir,0}$ 
with virial radius of a present subhalo $r_{\rm vir,0}$ and $r_{f}(z)=r_{f}/(1+z)$. Note that $\mv$ and $\rv$ in the right-hand side denote the virial mass and radius of a main 
progenitor at $z$, different from $M_{\rm vir,0}$ and $r_{\rm vir,0}$. We approximate the main progenitor mass at $z$ as $M_{\rm vir}(z)=D(z)M_{\rm vir,0}/D(z=0)$ where 
$D(z)$ is the linear growth factor. 

Figures \ref{fig:zr05}--\ref{fig:zr10} plot the analytical predictions of our density parity model for $\rth(z)/\rv(z)$ as gray areas for the cases of $\rf(z=0)/(\dunit)=0.5$ and $1$, 
respectively. As can be seen, the analytic predictions agree quite well with the numerically obtained  results for the transition radius threshold as a function 
of the progenitor redshifts in all of the four $\lmv$-intervals for both of the cases of $\rf(z=0)$.  This good agreement turns out to hold even when the analytical predictions are 
evaluated from different formulae for $c(M,z)$ \citep[e.g.,][]{CN07,lud-etal16} and the numerical results are obtained 
from the IllustrisTNG 300-1 hydrodynamical simulations which have different mass-concentration parameter relationship \citep[e.g.,][]{BB21}. 
Although the analytic evaluation of $\rth(z)$ resorted to the empirical relation of $c(M,z)$  and the crude approximation of the progenitor mass, the success of our density parity 
model and its robustness against the variation of $c(M,z)$ proves its validity, supporting its key assumption that the peculiar alignments of the subhalo inner spins with the Tweb 
major principal axes are generated when the internal tension is not high enough to resist the tidal compression. 

\section{Summary and Conclusion}\label{sec:sum}

Analyzing the subhalo catalog in the mass range of $11.8\le \log[\mv/(\munit)]\le 12.6$ at redshifts $0\le z\le 3$ from the IllustrisTNG 300-1 Dark 
simulations\citep{tngintro1, tngintro2, tngintro3, tngintro4, tngintro5, illustris19}, 
we have numerically determined the redshift evolution of the inner spin alignments of the subhalos with the Tweb principal axes and explored how well the 
inner spins of the subhalos observed at $z=0$ retain their memory of the high-$z$ Tweb principal axes measured at the locations of their progenitors. 
We have also constructed an analytic model, postulating that the radius-dependent transition of the subhalo inner spins occurs when the density contributed 
by the inner mass at the virial distance (a measure of the internal tension) is on a parity with that contributed by the virial mass at a distance equivalent to the smoothing 
scale of the Tweb (a measure of the tidal compression). 

\begin{figure}[ht]
\centering
\includegraphics[height=18cm,width=14cm]{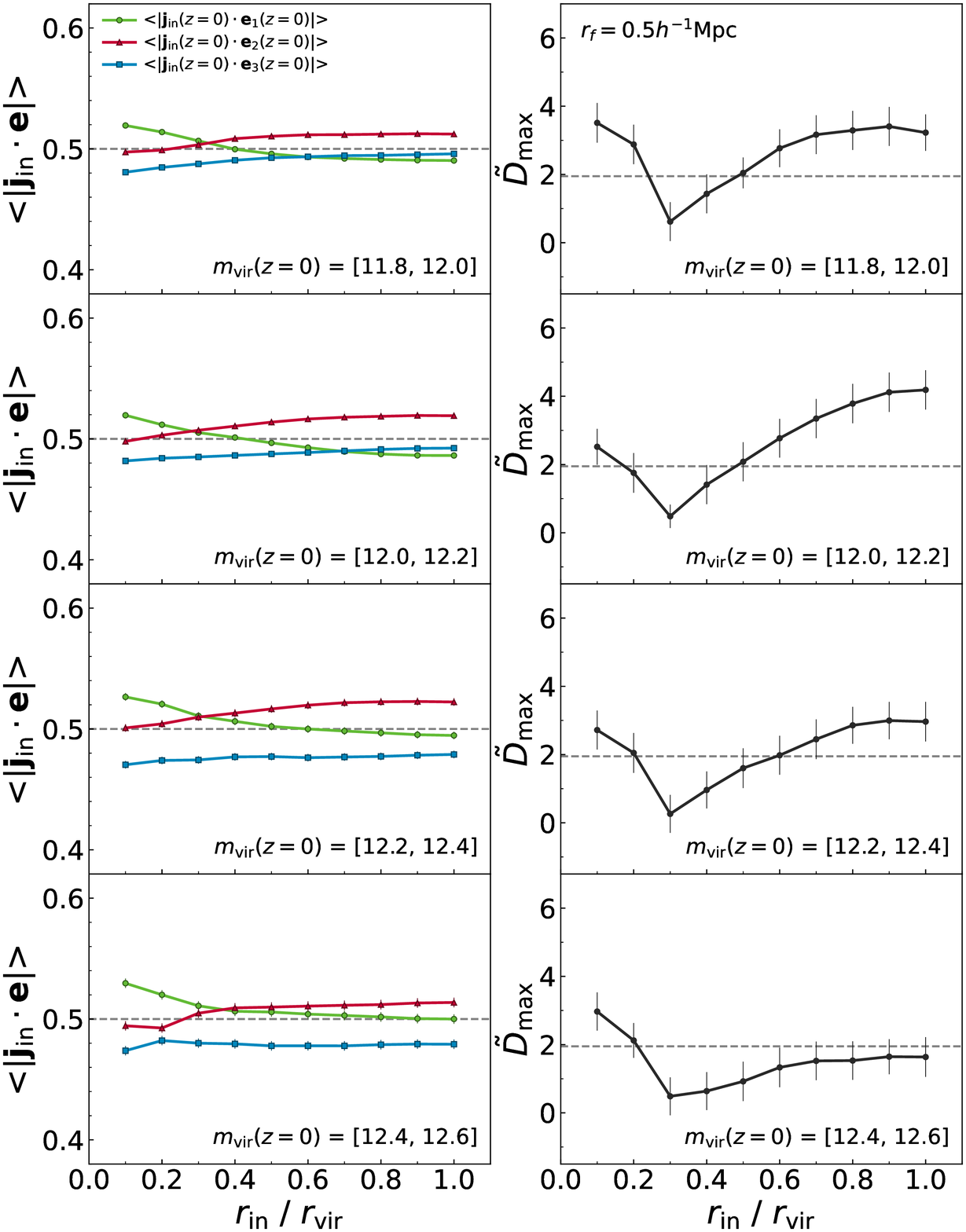}
\caption{(Left panels): alignments of the inner spins of the subhalos with the Tweb principal axes on the scale of $\rf/(\dunit)=0.5$ as a function of the inner-to-virial 
radius ratio for four different cases of the logarithmic mass intervals at $z=0$. (Right panels): maximum distance between $P(\langle\vert\jn(z)\cdot{\bf e}_{1}(z)\vert)$ 
and $P(\langle\vert\jn(z)\cdot{\bf e}_{2}(z)\vert)$ multiplied by $\sqrt{N_{\rm in}/2}$ as a function of the inner-to-virial radius ratio from the four $\lmv$-intervals. 
The errorbars in all panels are computed via the Jackknife resampling method.} 
\label{fig:cl005}
\end{figure}

\begin{figure}[ht]
\centering
\includegraphics[height=18cm,width=14cm]{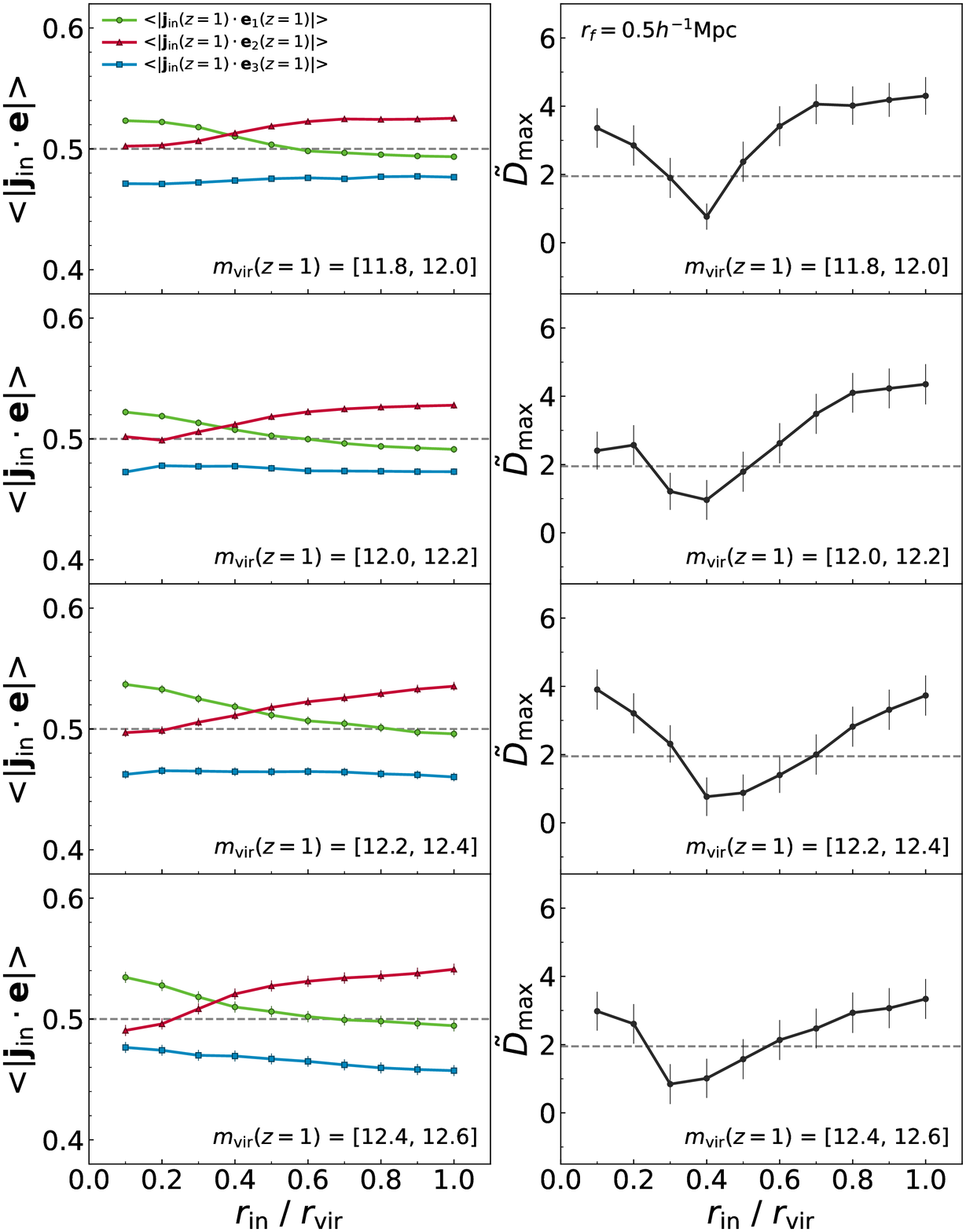}
\caption{Same as Figure \ref{fig:cl005} but at $z=1$.}
\label{fig:cl105}
\end{figure}

\begin{figure}[ht]
\centering
\includegraphics[height=18cm,width=14cm]{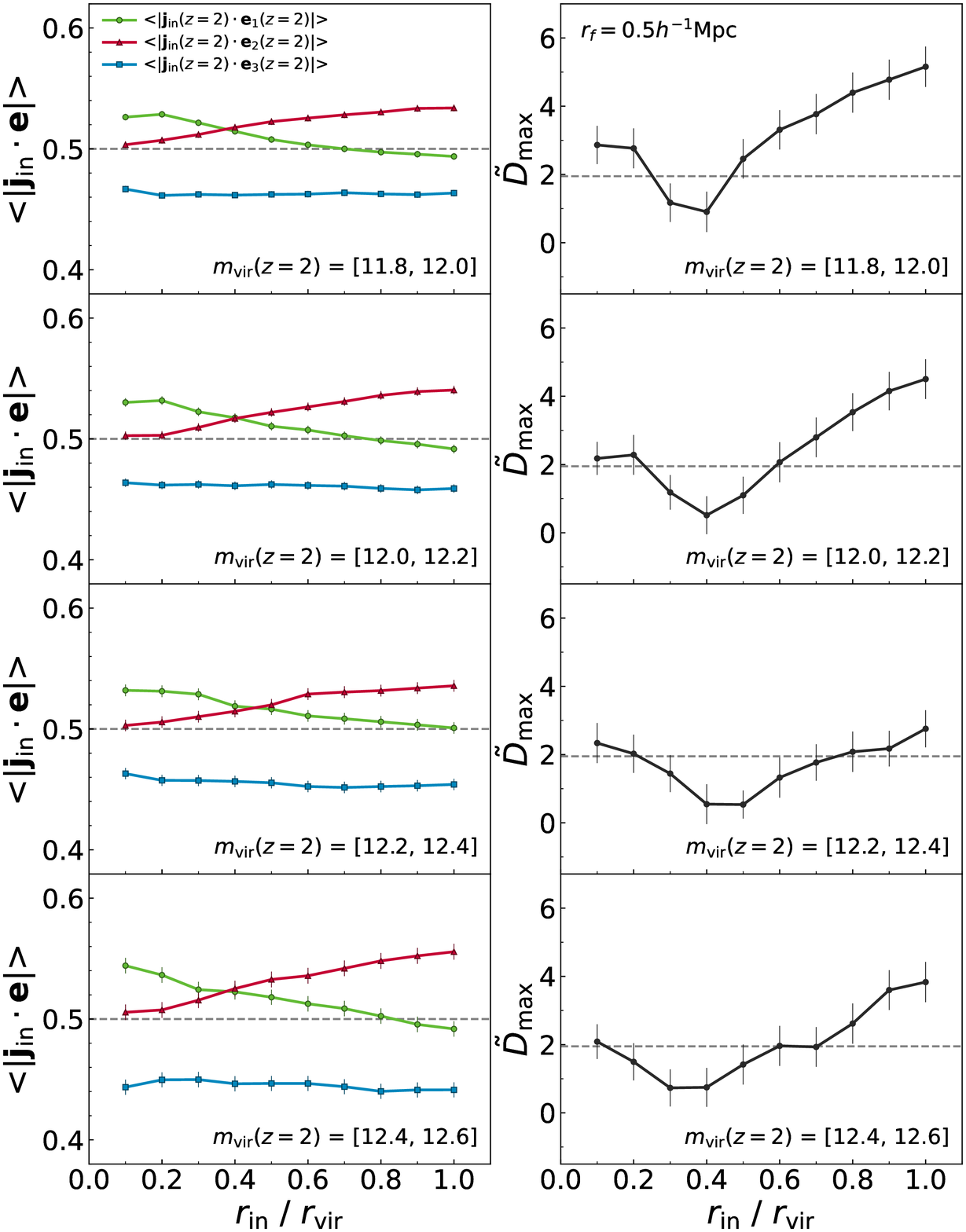}
\caption{Same as Figure \ref{fig:cl005} but at $z=2$.}
\label{fig:cl205}
\end{figure}

\begin{figure}[ht]
\centering
\includegraphics[height=18cm,width=14cm]{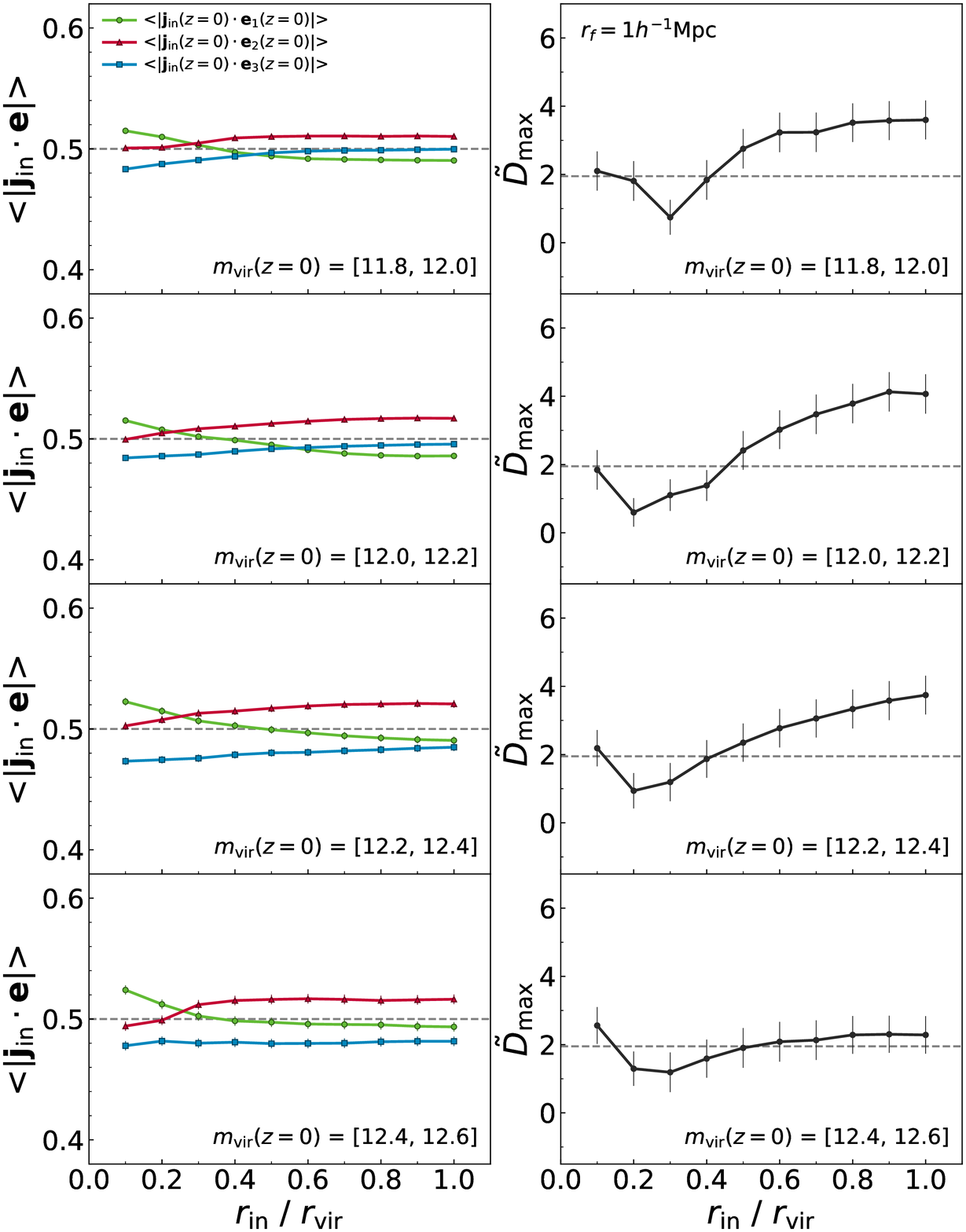}
\caption{Same as Figure \ref{fig:cl005} but on the scale of $\rf/(\dunit)=1$.}
\label{fig:cl010}
\end{figure}

\begin{figure}[ht]
\centering
\includegraphics[height=18cm,width=14cm]{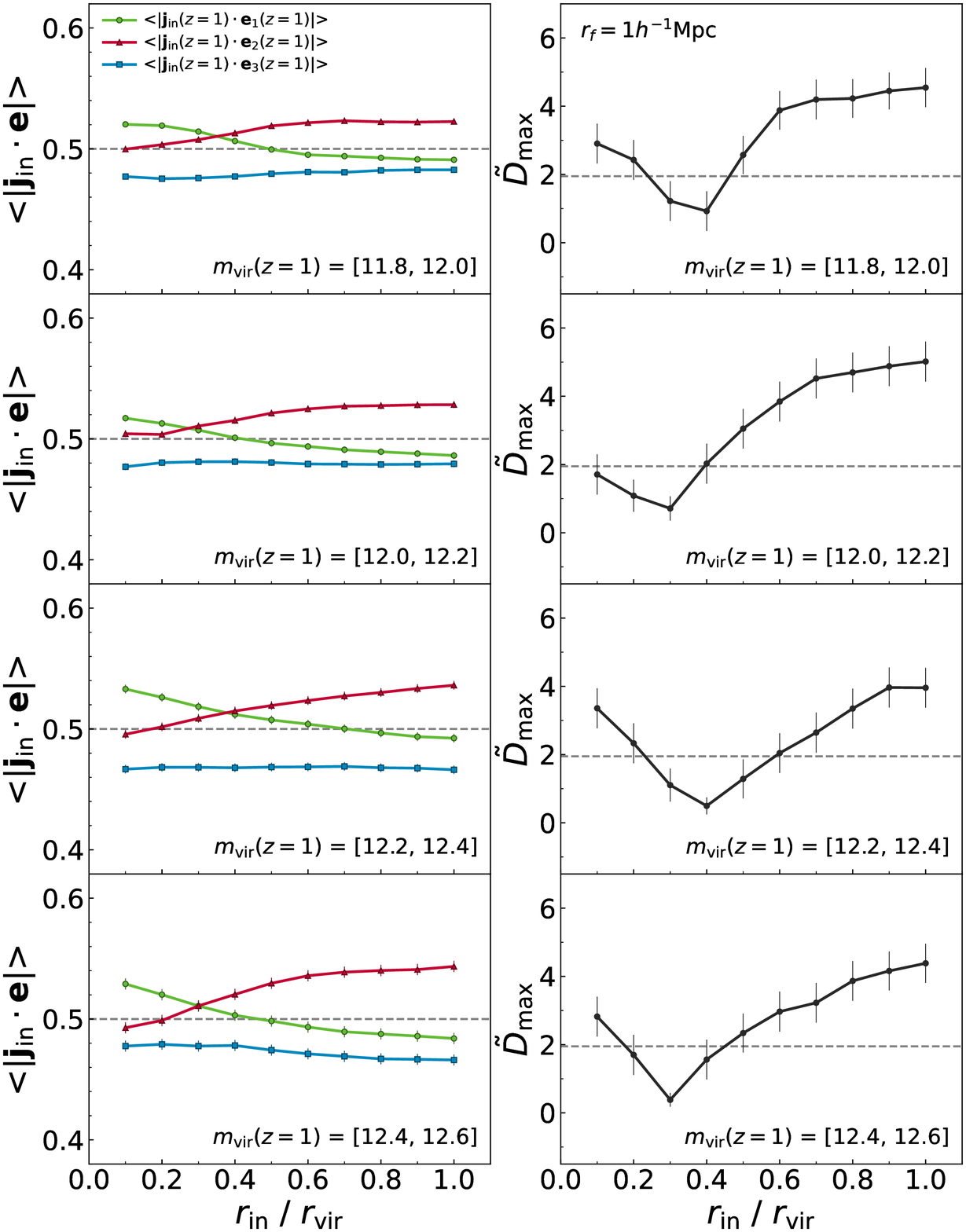}
\caption{Same as Figure \ref{fig:cl010} but at $z=1$.}
\label{fig:cl110}
\end{figure}

\begin{figure}[ht]
\centering
\includegraphics[height=18cm,width=14cm]{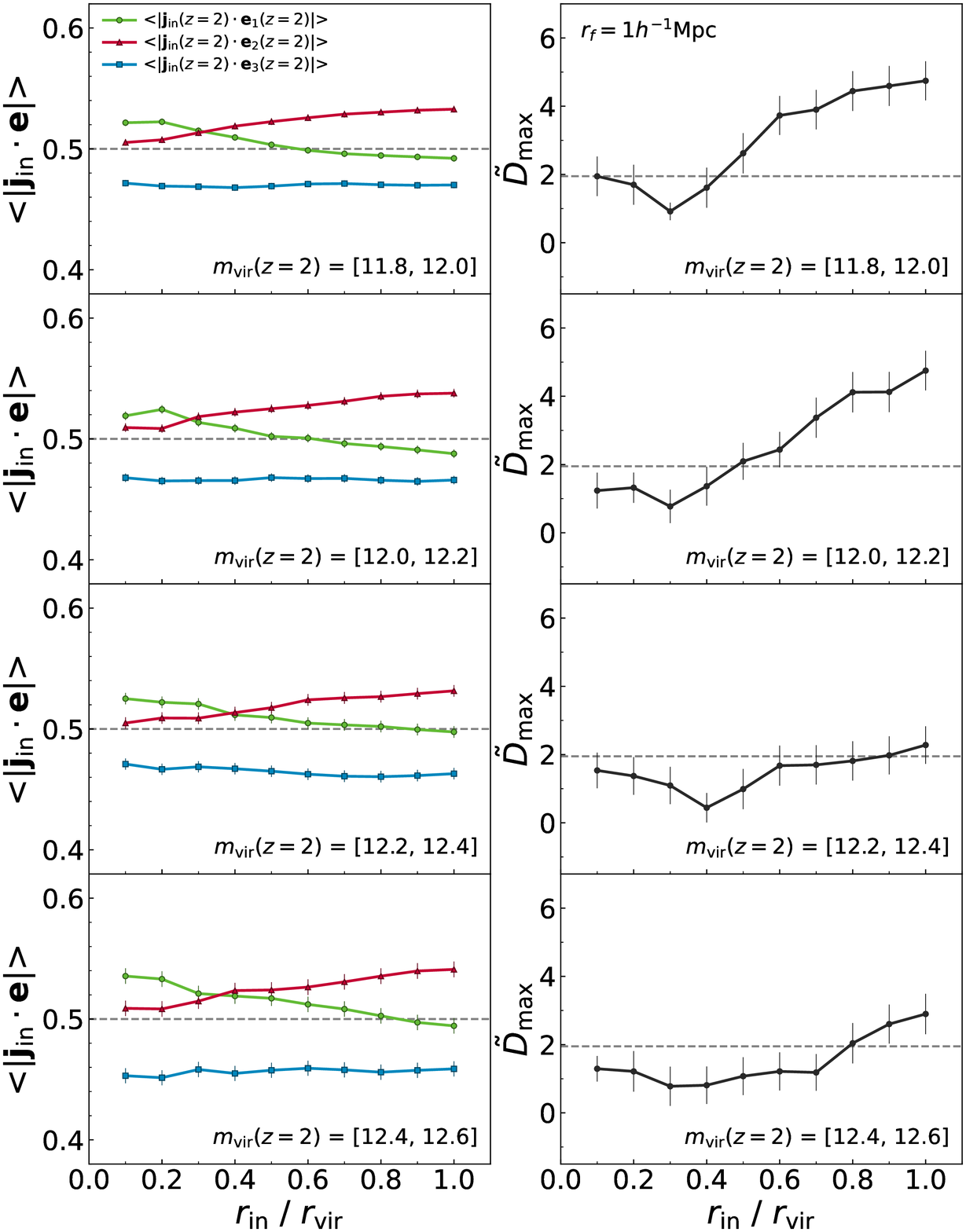}
\caption{Same as Figure \ref{fig:cl010} but at $z=2$.}
\label{fig:cl210}
\end{figure}

\begin{figure}[ht]
\centering
\includegraphics[height=14cm,width=14cm]{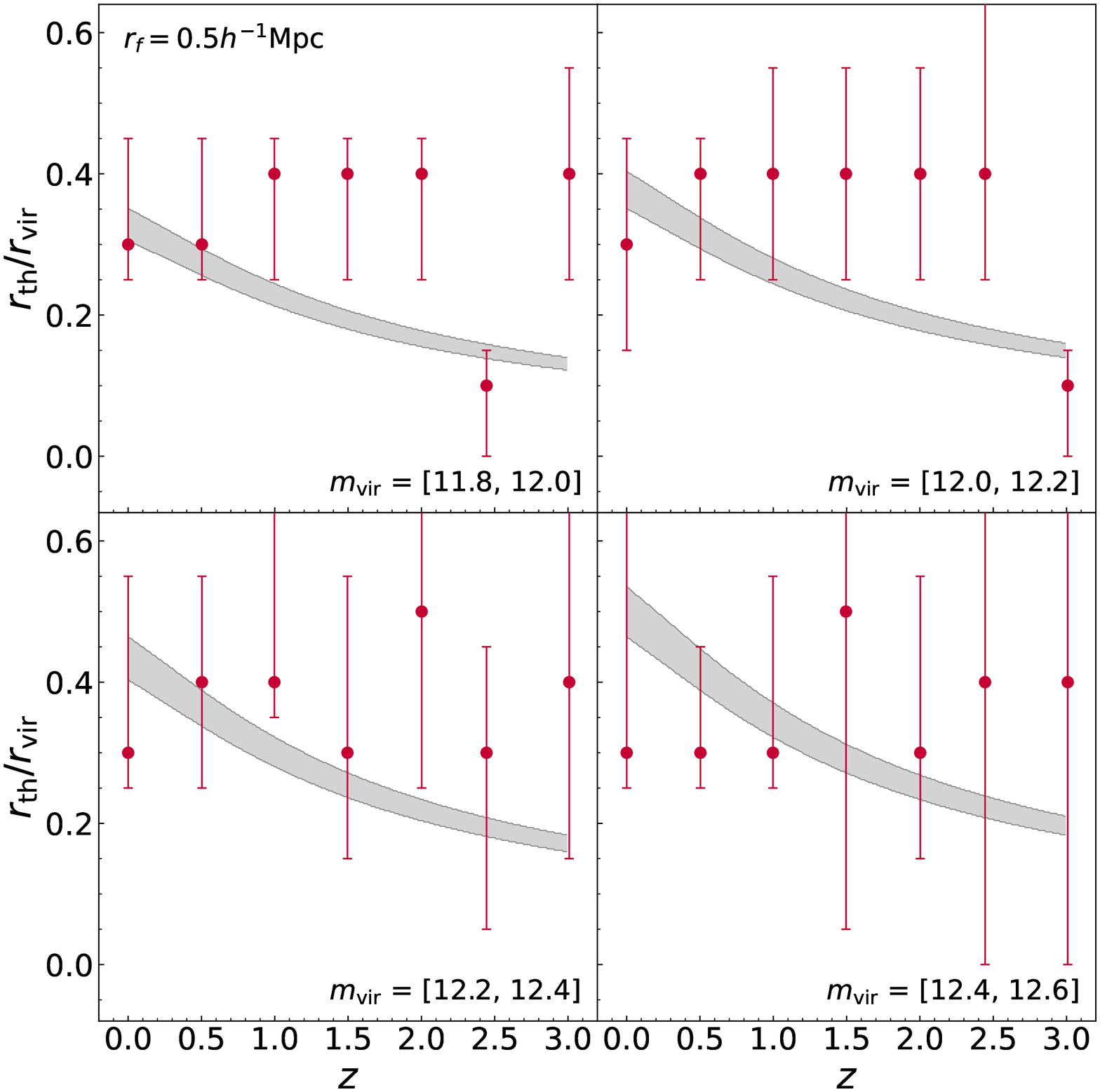}
\caption{Numerically obtained transition threshold radius rescaled by the virial boundary (red filled circles) compared with 
the analytic predictions (gray areas) as a function of redshift in the four different ranges 
of the logarithmic total masses of the subhalos for the case of $\rf/(\dunit)=0.5$.} 
\label{fig:z05}
\end{figure}

\begin{figure}[ht]
\centering
\includegraphics[height=14cm,width=14cm]{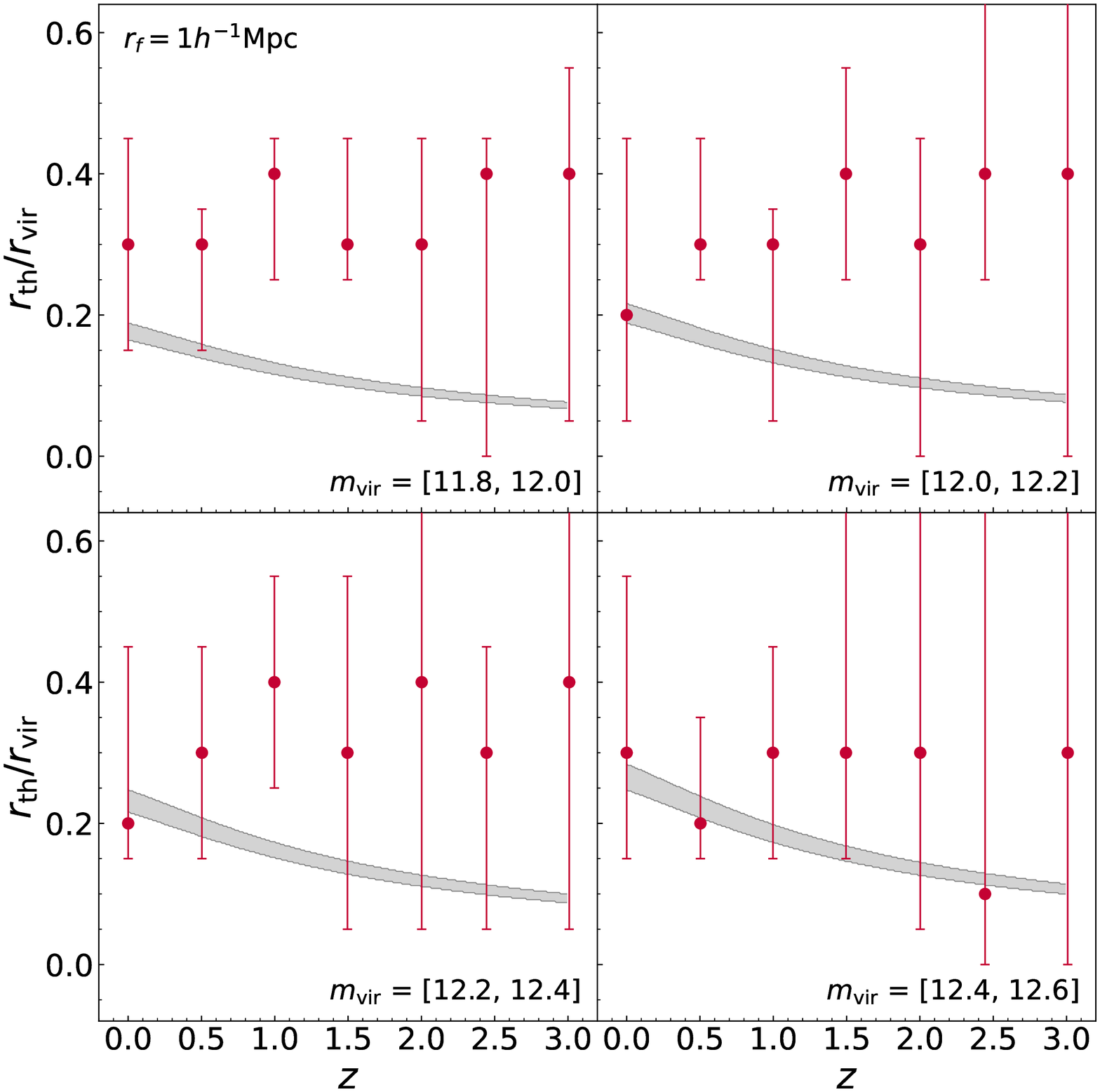}
\caption{Same as Figure \ref{fig:z05} but for the case of $\rf/(\dunit)=1$.}
\label{fig:z10}
\end{figure}

\begin{figure}[ht]
\centering
\includegraphics[height=18cm,width=14cm]{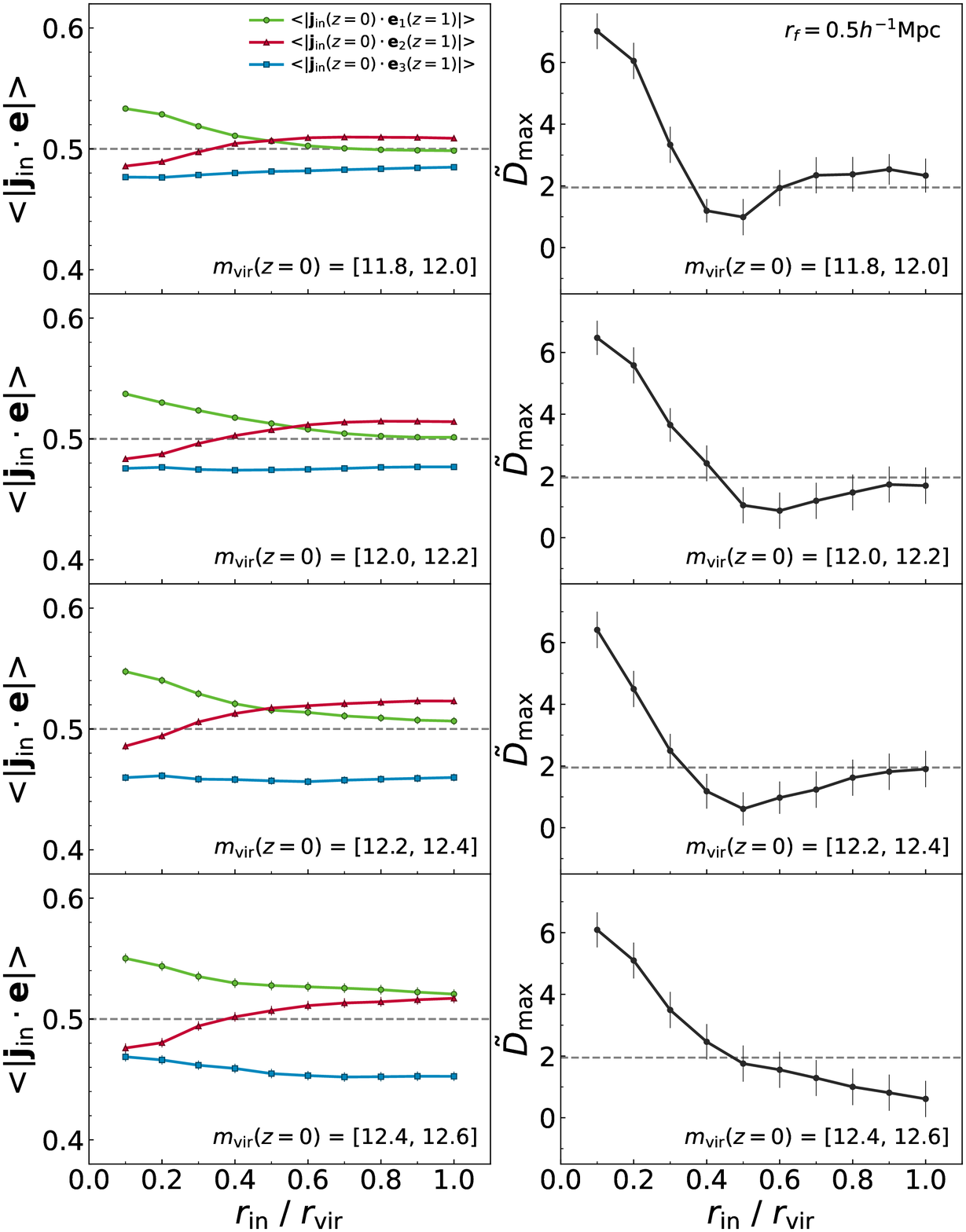}
\caption{Same as Figure \ref{fig:cl105} but for the case that the subhalo inner spins are measured at $z=0$ 
while the Tweb principal axes are measured at the progenitor redshifts, $z=1$.}
\label{fig:cr105}
\end{figure}

\begin{figure}[ht]
\centering
\includegraphics[height=18cm,width=14cm]{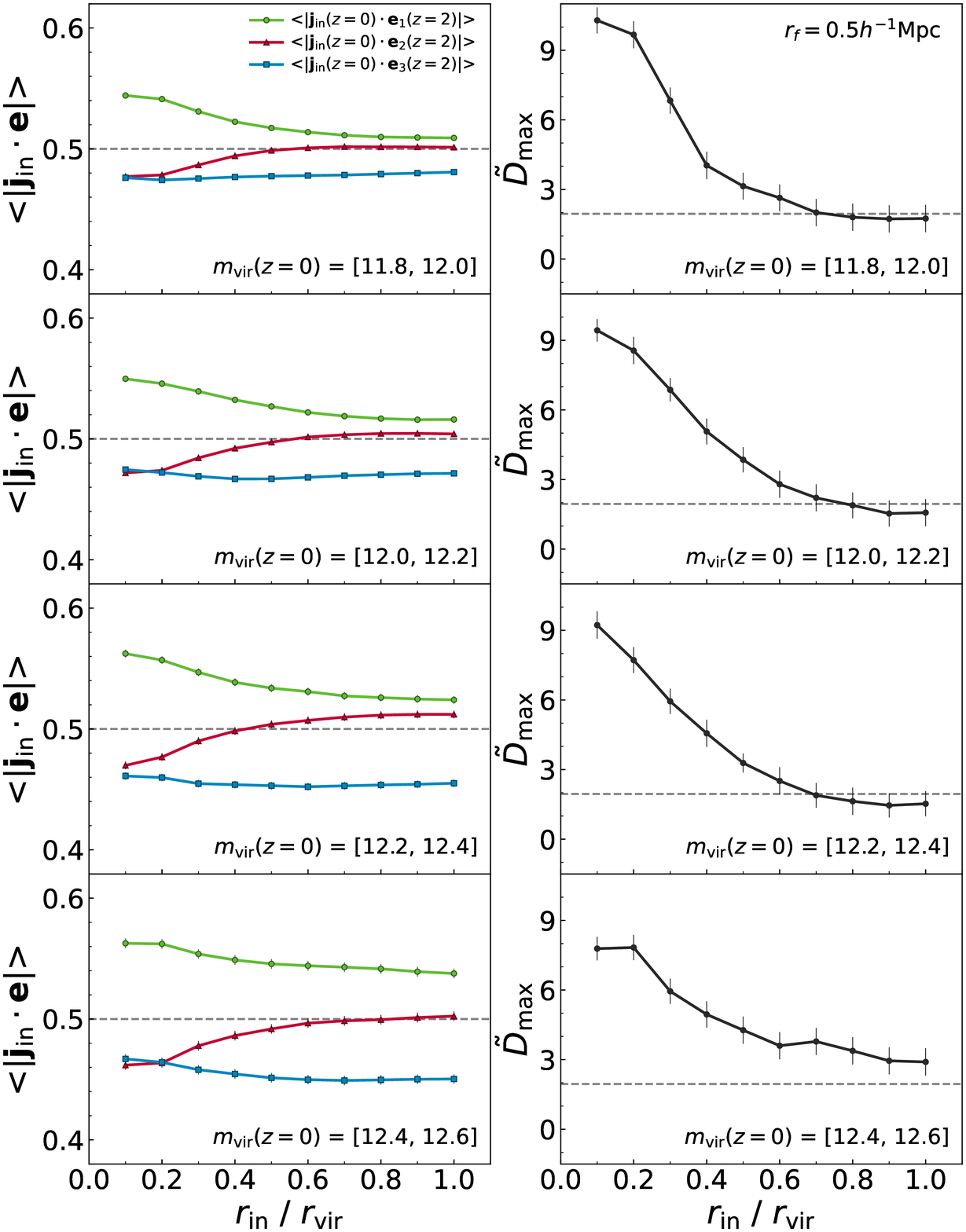}
\caption{Same as Figure \ref{fig:cl205} but for the case that the subhalo inner spins are measured at $z=0$ 
while the Tweb principal axes are measured at the progenitor redshifts, $z=2$.} 
\label{fig:cr205}
\end{figure}

\begin{figure}[ht]
\centering
\includegraphics[height=18cm,width=14cm]{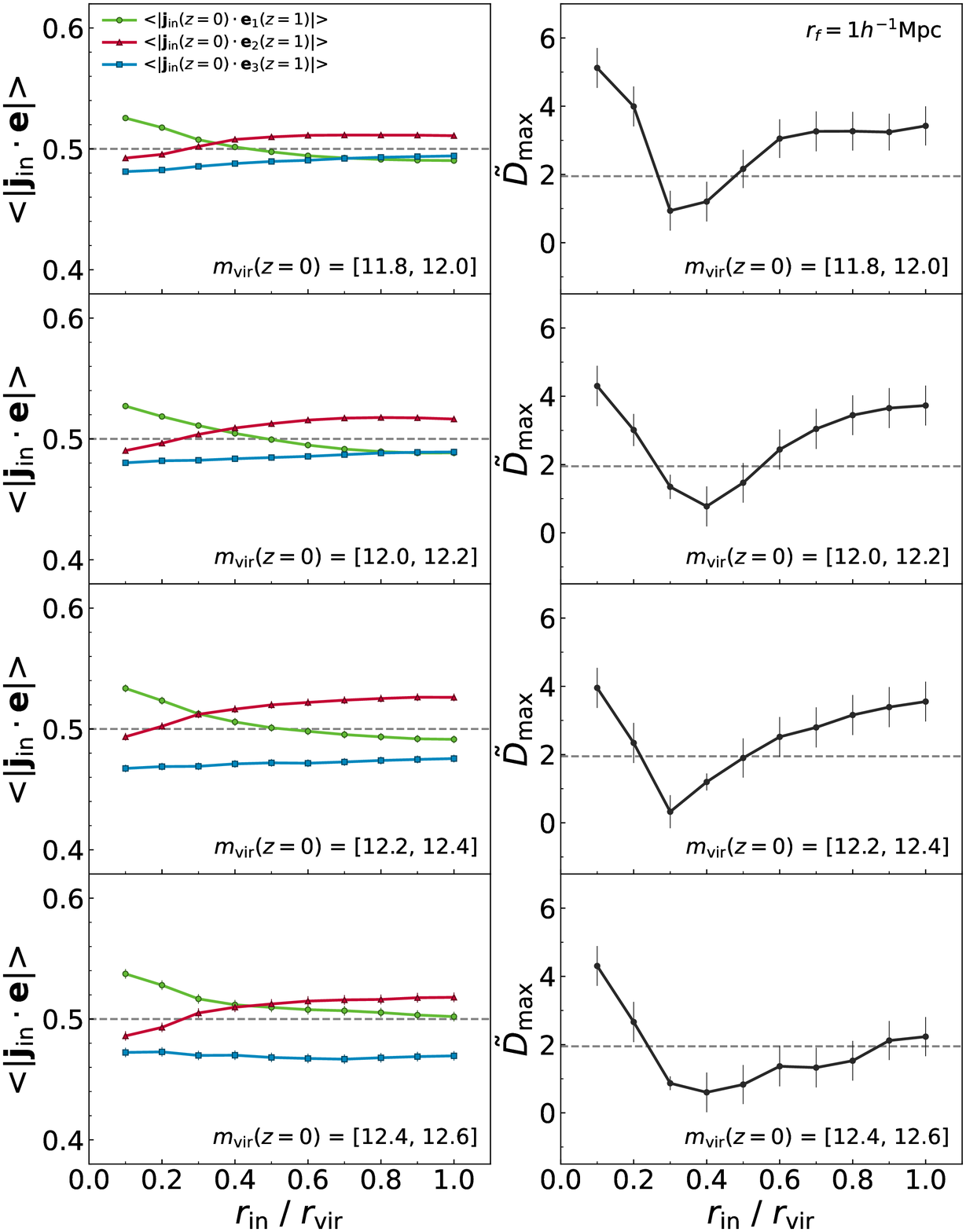}
\caption{Same as Figure \ref{fig:cr105} but for the case of $\rf/(\dunit)=1$.}
\label{fig:cr110}
\end{figure}

\begin{figure}[ht]
\centering
\includegraphics[height=18cm,width=14cm]{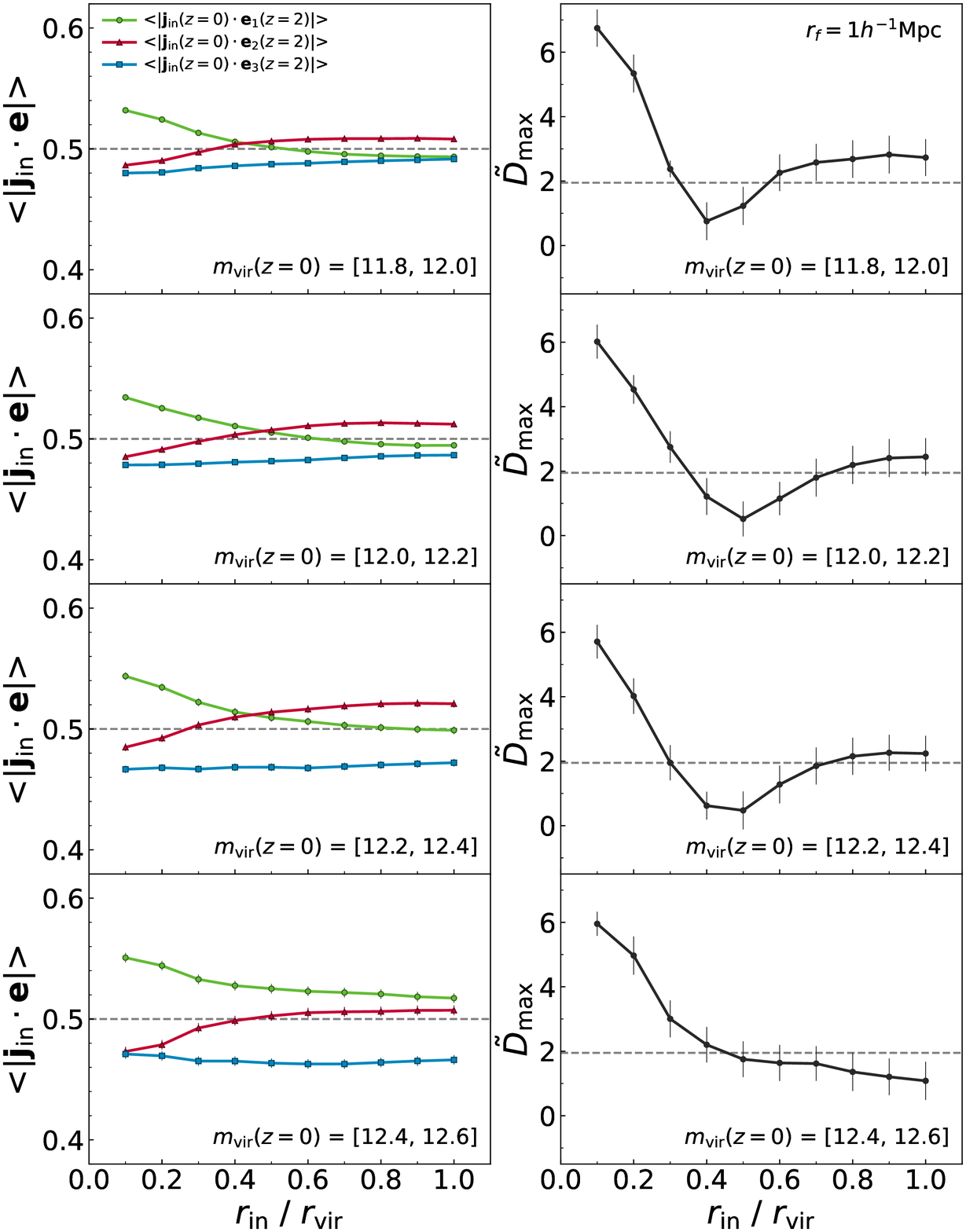}
\caption{Same as Figure \ref{fig:cr205} but for the case of $\rf/(\dunit)=1$.}
\label{fig:cr210}
\end{figure}

\begin{figure}[ht]
\centering
\includegraphics[height=14cm,width=14cm]{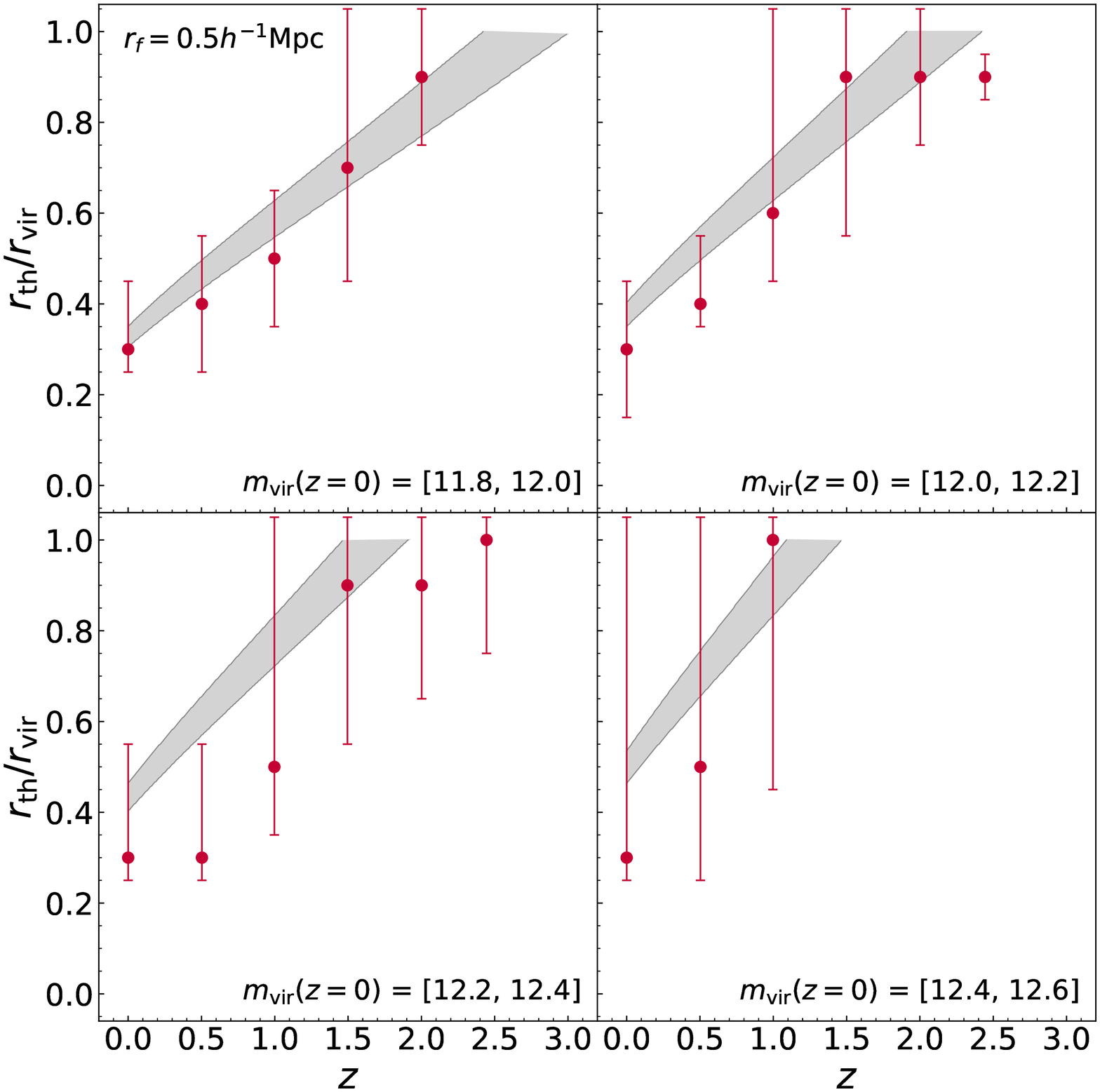}
\caption{Same as Figure \ref{fig:z05} but for the case that the subhalo inner spins are measured at the present epoch 
while the Tweb principal axes are measured at the progenitor redshifts, $z$.}
\label{fig:zr05}
\end{figure}

\begin{figure}[ht]
\centering
\includegraphics[height=14cm,width=14cm]{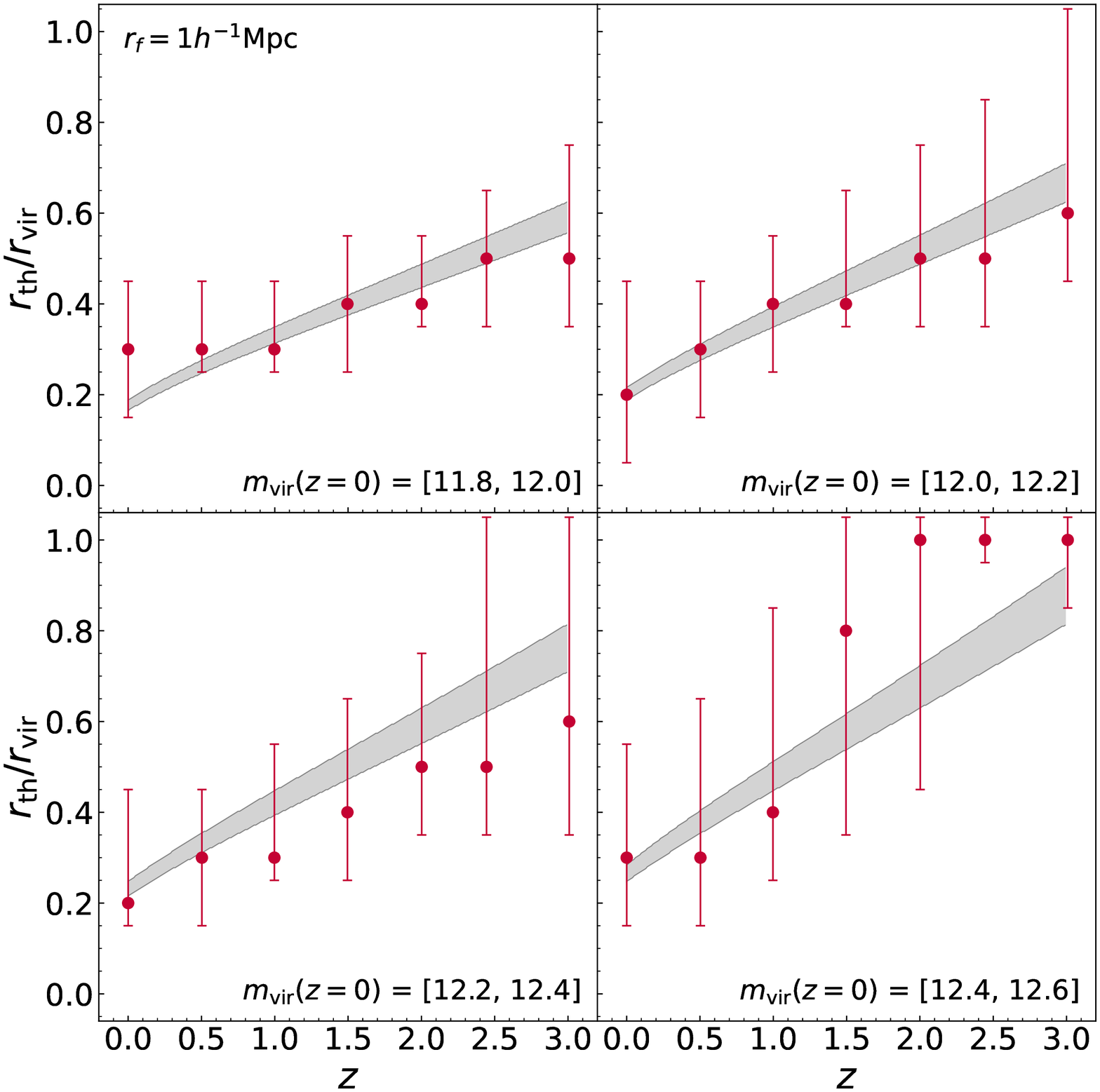}
\caption{Same as Figure \ref{fig:zr05} but for the case of $\rf/(\dunit)=1$.}
\label{fig:zr10}
\end{figure}

\begin{figure}[ht]
\centering
\includegraphics[height=16cm,width=14cm]{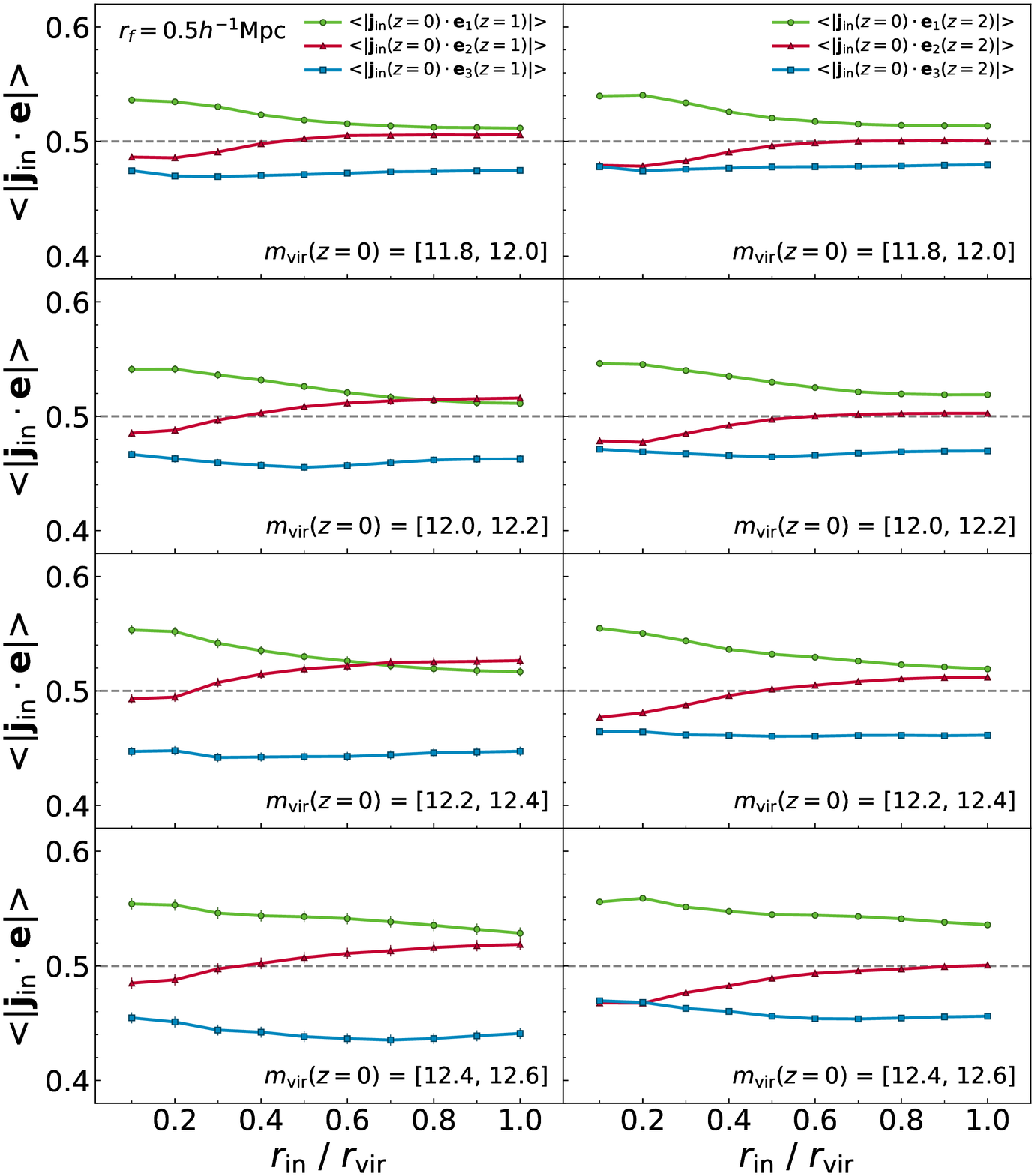}
\caption{(Left panel): same as the left panel of Figure \ref{fig:cr105} but for the case that the high-$z$ Tweb is 
determined at the locations of the non-main (less massive) progenitors of the present subhalos than the main counterparts. 
(Right panel): same as the left panel but at $z=2$.}
\label{fig:cr05_non}
\end{figure}

\begin{figure}[ht]
\centering
\includegraphics[height=16cm,width=14cm]{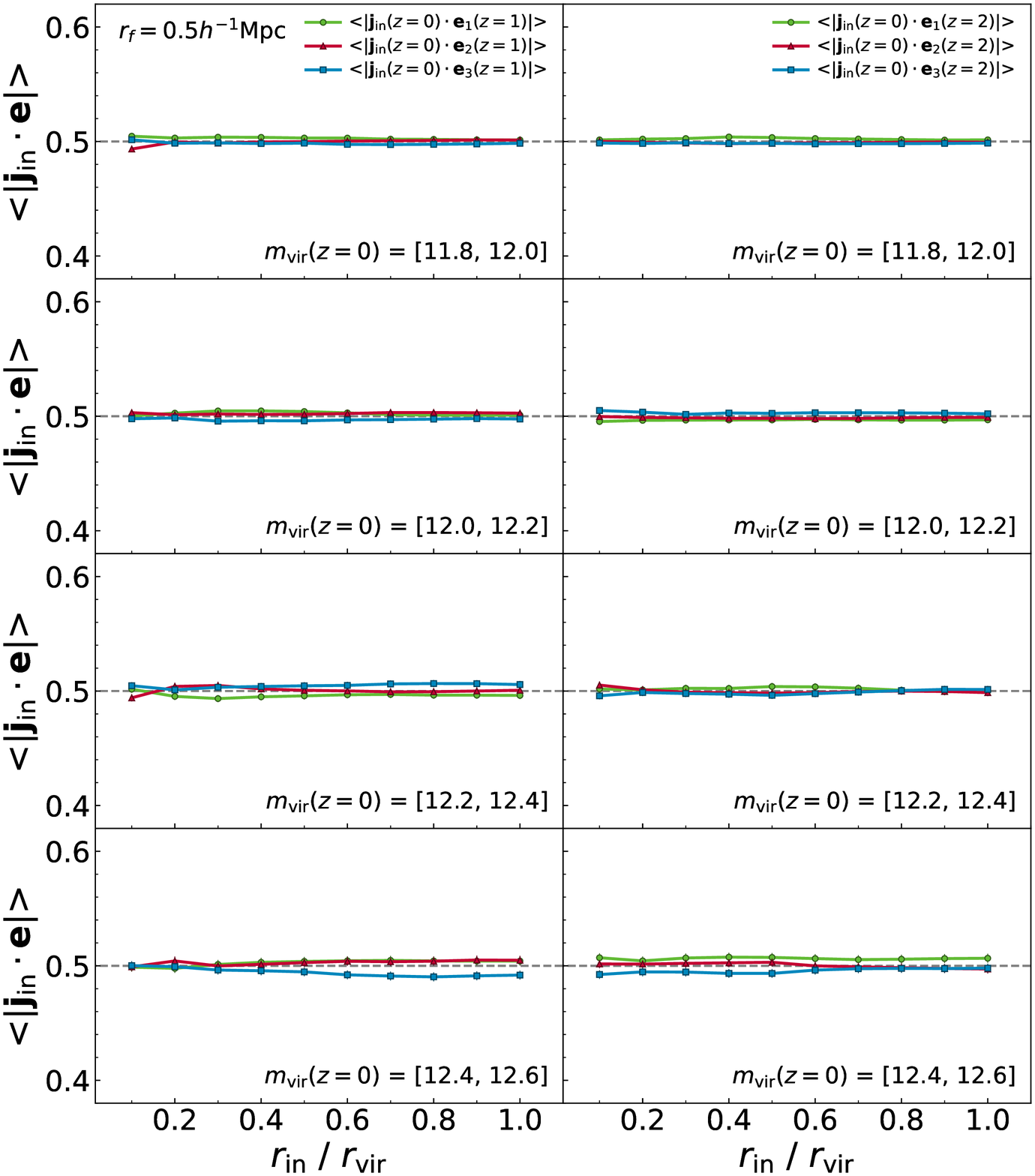}
\caption{(Left panel): same as the left panel of Figure \ref{fig:cr105} but for the case that the high-$z$ Tweb is 
determined at the randomly shuffled locations of the main progenitors. (Right panel): same as the left panel but at $z=2$.}
\label{fig:cr05_ran}
\end{figure}

The main results are summarized in the following. 
\begin{itemize}
\item
The threshold radii for the inner spin transitions between the Tweb intermediate and major principal axes do not show strong variation with redshifts 
(Figures \ref{fig:cl005}--\ref{fig:z10}). 
\item
The Tweb major principal axes at the progenitor locations are more strongly aligned with the inner spins of the descendant subhalos at the present epochs 
than those of the progenitors at the same redshifts. This alignment becomes stronger as the redshift gap between the descendants and the progenitors 
becomes larger. 
\item
The threshold radii for the present spin transitions between the high-$z$ Tweb intermediate and major principal axes at the progenitor locations strongly vary with 
redshifts, having much larger values at each redshift than those of the high-$z$ progenitor spins, regardless of the subhalo mass (Figures \ref{fig:cr105}--\ref{fig:zr10}). 
\item
The analytic model describes quite well the numerically obtained redshift evolution of the transition threshold radii for all of the cases of subhalo mass and smoothing scales 
considered (Figures \ref{fig:z05}--\ref{fig:z10} and \ref{fig:zr05}--\ref{fig:zr10}). 
\end{itemize}

Given that our results have been obtained from the subhalos on the galactic mass scale and that the subhalo stellar spins have the same alignment tendency 
as the subhalo DM inner spins at the innermost radii as shown by \citet{ML22}, a crucial implication of our results is that the stellar spins of the present 
galaxies must be a good indicator of the high-$z$ tidal field. This implication in turn hints at the possibility of obtaining information on the early density field from the present galaxy 
spin field since at high redshifts the density field can be linearly constructed from the curl-free tidal field \citep{lib-etal13}. In all of the previous theoretical works that attempted 
to connect the present subhalo spin field to the initial conditions, it was always implicitly assumed that the alignment tendency of the observable stellar spins with the Tweb 
principal axes must be the same as or at least similar to that of the subhalo virial spins 
\citep[e.g.,][]{LP00,LP01,yu-etal19,lee-etal20,LL20,yu-etal20,mot-etal21,mot-etal22}.  
While in our prior works \citep{ML22} this conventional assumption was disproved \citep{lee-etal21,LM22}, in the current work a true bridge has been found between the 
observable galaxy stellar spins and the high-$z$ universe. Our future work is in the direction of probing the high-$z$ density field with this bridge.

\acknowledgments

We thank an anonymous referee for his/her very helpful comments which helped us greatly improve the original manuscript. 
The IllustrisTNG simulations were undertaken with compute time awarded by the Gauss Centre for Supercomputing (GCS) 
under GCS Large-Scale Projects GCS-ILLU and GCS-DWAR on the GCS share of the supercomputer Hazel Hen at the High 
Performance Computing Center Stuttgart (HLRS), as well as on the machines of the Max Planck Computing and Data Facility 
(MPCDF) in Garching, Germany. 
JSM acknowledges the support by the National Research Foundation (NRF) of Korea grant funded by the Korean government (MEST) (No. 2019R1A6A1A10073437).
JL acknowledges the support by Basic Science Research Program through the NRF of Korea funded by the Ministry of Education (No.2019R1A2C1083855).

\clearpage
\clearpage
\clearpage
\clearpage
\clearpage
\clearpage
\clearpage
\clearpage
\clearpage
\clearpage
\clearpage
\clearpage
\clearpage
\clearpage
\clearpage
\clearpage
\end{document}